\documentclass[11pt]{article}
\usepackage{blois}
\usepackage{amsmath}
\usepackage{amssymb}
\usepackage{amsthm}
\usepackage{amsfonts}

\newcommand{\m}{M_{H^{\pm}}}
\newcommand{\g}{\,\mbox{GeV}}

\newcommand{\lc}{\lambda_3}

\newcommand{\lczp}{\lambda_{345}}
\newcommand{\rg}{R_{\gamma\gamma}}

\newcommand{\fr}{\frac}
\newcommand{\relic}{\Omega_{DM}h^2}

\def\be{\begin{equation}}
\def\ee{\end{equation}}
\def\bea{\begin{eqnarray}}
\def\eea{\end{eqnarray}}

\newcommand{\Photo}{}

\begin{document}
\vspace*{4cm}
\title{DARK MATTER AND 125 GEV HIGGS FOR IDM}

\author{ D. SOKO{\L}OWSKA }

\address{University of Warsaw, Faculty of Physics,\\Ho\.{z}a 69, 00-681 Warsaw, Poland}

\maketitle\abstracts{
We discuss a scalar Dark Matter candidate from the Inert Doublet Model in
light of discovery of a 125 GeV SM-like Higgs boson at the LHC.
%We take into account the vacuum stability, perturbative unitarity, 
%electroweak precision tests and the condition for existence of the true
%(and not metastable) inert vacuum. 
We explore the possibility of using the recent and future data from LHC 
experiments, namely the Higgs diphoton decay measurements, to constrain the properties of Dark Matter particles.
%independently on direct and indirect DM detection experiments.
}

\section{Introduction}
According to the standard cosmological model around 25\% of the Universe is made of cold, non-baryonic, neutral and very weakly interacting particles. 
%Various particle physics' models can provide such a state, which is usually stable due to a certain discrete symmetry.
The Inert Doublet Model (IDM) is one of the simplest extensions of the Standard Model (SM) that can provide such Dark Matter (DM) candidate. The scalar sector in the IDM is extended with respect to the SM-like Higgs doublet $\Phi_S$ by a second $SU(2)$ doublet, $\Phi_D$, which is odd under a $D\;(Z_2)$ symmetry: $\Phi_S\to \Phi_S, \Phi_D \to - \Phi_D, \textrm{SM fields} \to \textrm{SM fields}$.\cite{Cao:2007rm,Barbieri:2006dq} 

The IDM can provide a viable DM candidate in agreement with collider constraints and relic density measurements in three regions of DM mass: $M_{DM} \lesssim 10 \g$, $40 \g \lesssim M_{DM} \lesssim 160 \g$ and $M_{DM} \gtrsim 500 \g$.\cite{Dolle:2009fn} Further constraints for the DM candidate can come from direct and indirect detection experiments. However, as for now there is no agreement how to consistently interpret various reported signals and the exclusion limits.\cite{Bergstrom:2012fi}

In this work we set constraints on scalar DM from the IDM by using the LHC Higgs data and WMAP relic density measurements. Combining the $h \to \gamma \gamma$ data for the SM-like Higgs with the WMAP results excludes a large part of the IDM parameter space, setting limits on DM that are stronger or comparable to these obtained by the DM detection experiments.

\section{The Inert Doublet Model}

The IDM is defined as a 2HDM with a $D$-symmetric potential and vacuum state:
\begin{equation}\begin{array}{c}
V=-\fr{1}{2}\left[m_{11}^2(\Phi_S^\dagger\Phi_S)\!+\! m_{22}^2(\Phi_D^\dagger\Phi_D)\right]+
\fr{\lambda_1}{2}(\Phi_S^\dagger\Phi_S)^2\! 
+\!\fr{\lambda_2}{2}(\Phi_D^\dagger\Phi_D)^2\\[2mm]+\!\lambda_3(\Phi_S^\dagger\Phi_S)(\Phi_D^\dagger\Phi_D)\!
\!+\!\lambda_4(\Phi_S^\dagger\Phi_D)(\Phi_D^\dagger\Phi_S) +\fr{\lambda_5}{2}\left[(\Phi_S^\dagger\Phi_D)^2\!
+\!(\Phi_D^\dagger\Phi_S)^2\right],\\[2mm]
\langle\Phi_S\rangle =\frac{1}{\sqrt{2}} \begin{pmatrix}0\\ v\end{pmatrix}\,,\qquad \langle\Phi_D\rangle = \frac{1}{\sqrt{2}}
\begin{pmatrix} 0 \\ 0  \end{pmatrix}, \quad v = 246 \textrm{ GeV},
\end{array}\label{dekomp_pol}\end{equation}
%with all  parameters real . The vacuum state in the IDM is given by: 
%\begin{equation}
%\langle\Phi_S\rangle =\frac{1}{\sqrt{2}} \begin{pmatrix}0\\ v\end{pmatrix}\,,\qquad \langle\Phi_D\rangle = \frac{1}{\sqrt{2}}
%\begin{pmatrix} 0 \\ 0  \end{pmatrix}, \quad v = 246 \textrm{ GeV}.\label{dekomp_pol}
%\end{equation}
and Yukawa interaction set to Model I.\cite{Cao:2007rm,Barbieri:2006dq}

In the IDM only one doublet, $\Phi_S$, is involved in the EW symmetry breaking. It provides a SM-like Higgs boson $h$, which has tree-level couplings to fermions and gauge bosons like in the SM with possible deviation from the SM in loop couplings. The second doublet, $\Phi_D$, is inert and contains four dark scalars $H, A, H^\pm$, that have no couplings to fermions. The lightest particle coming from this doublet is stable, being a good DM candidate. 

The IDM can be described by the masses of scalar particles and their physical couplings: $\lambda_{345}=\lambda_3 + \lambda_4 + \lambda_5$ is related to $hHH$ and $hhHH$ vertices, while $\lambda_3$ gives $h H^+ H^-$ and $\lambda_2$: $HHHH$.
Parameters of the IDM are constrained by various theoretical and experimental conditions. In our analysis we use vacuum stability constraints, that ensure the potential is bounded from below. We also demand, that the state (\ref{dekomp_pol}) is the global, and not just a local minimum.\cite{Krawczyk:2010,Sokolowska:2011aa} Parameters of the potential should also fulfill perturbative unitarity bounds.\cite{Kanemura:1993}

The value of the Higgs boson mass, $M_h = 125 \g$, and above conditions provide the following constraints for the parameters of the potential:
$\lambda_1 = 0.258, \; m_{22}^2\lesssim 9\cdot10^4\g^2, \;  \lambda_3, \lambda_{345} > -\sqrt{\lambda_1\lambda_2} \geqslant -1.47, \; \lambda_{2}^{\textrm{max}} = 8.38$.\cite{Swiezewska:2012}

Masses of dark particles are constrained by the LEP measurements and EWPT to be:
$\m +M_H>M_{W},\  \m+ M_A>M_W,\ M_H+M_A >M_Z,\ 2\m > M_Z,\ \m>70\g$
with an excluded region where simultaneously $M_H< 80\g,\  M_A< 100\g\ \textrm{and}\  M_A - M_H> 8\g$.\cite{Gustafsson:2009}

\section{The diphoton decay rate $\rg$ in the IDM}

%A SM-like Higgs particle was discovered at the LHC in 2012. 
%Departure from the SM Higgs is given by 
$\rg$ is the ratio of the diphoton decay rate of the Higgs particle $h$ observed at the LHC to the SM prediction.
%and it is sensitive to the "new physics". 
%LHC results suggest that signal strengths of the Higgs boson decaying into two photons can differ from the value expected in the Standard Model. 
The current measured values of $\rg$ 
%provided by the ATLAS and the CMS collaborations 
are $\rg = 1.65\pm0.24\mathrm{(stat)}^{+0.25}_{-0.18}\mathrm{(syst)}$ for ATLAS and $\rg = 0.79^{+0.28}_{-0.26}$ for CMS.\cite{ATLAS:2013oma,CMStalk}
%\begin{eqnarray}
%\textrm{ATLAS} &: & \rg = 1.65\pm0.24\mathrm{(stat)}^{+0.25}_{-0.18}\mathrm{(syst)}, \label{rg_atlas} \\
%\textrm{CMS} & : &  \rg = 0.79^{+0.28}_{-0.26}. \label{rg_cms}
%\end{eqnarray}
%\begin{equation}
%\textrm{ATLAS} \; : \;  \rg = 1.65\pm0.24\mathrm{(stat)}^{+0.25}_{-0.18}\mathrm{(syst)}, \quad
%\textrm{CMS} \; : \;   \rg = 0.79^{+0.28}_{-0.26}. \label{rg_cmsatlas}
%\end{equation}
They are in 2$\sigma$ agreement with the SM value $\rg = 1$, however a deviation from that value is still possible and would be an indication of physics beyond the SM.
 
The ratio $\rg$ in the IDM is given by:
\begin{equation}\label{rgg}
R_{\gamma \gamma}:=\frac{\sigma(pp\to h\to \gamma\gamma)^{\textrm{IDM}}}{\sigma(pp\to h\to \gamma\gamma)^{\textrm  {SM}}}
%\approx\frac{\textrm{Br}(h\to\gamma\gamma)^{\textrm {IDM}}}{\textrm{Br}(h\to\gamma\gamma)^{\textrm {SM}}},
\approx \frac{\Gamma(h\to \gamma\gamma)^{\mathrm{IDM}}}{\Gamma(h\to \gamma\gamma)^{\mathrm{SM}}}\frac{\Gamma(h)^{\mathrm{SM}}}{\Gamma(h)^{\mathrm{IDM}}} \, ,
\end{equation}
where $\Gamma(h)^{\mathrm{SM}/\mathrm{IDM}}$ are the total decay widths of $h$ in the SM and the IDM, while $\Gamma(h\to \gamma\gamma)^{\mathrm{SM}/\mathrm{IDM}}$ are the respective partial decay widths for $h\to\gamma\gamma$. 
%In (\ref{rgg}) the facts that the main production channel is gluon fusion and that the Higgs particle from the IDM is SM-like, so $\sigma(gg\to h)^{\textrm{IDM}} = \sigma(gg\to h)^{\textrm{SM}}$, were used. 
In the IDM two sources of deviation from $\rg=1$ are possible.
%\begin{itemize} 
%\item 
First is a $H^\pm$ contribution to the partial decay width:\cite{Djouadi:2005,Arhrib:2012}
%$$
%\Gamma(h\to\gamma\gamma)^{\textrm{IDM}}=\frac{G_F\alpha^2M_h^3}{128\sqrt{2}\pi^3}\left | \mathcal{M}^{\textrm{SM}}+\delta\mathcal{M}^{\textrm{IDM}}\right |^2,
%$$
%\begin{equation}
% \Gamma(h \rightarrow \gamma\gamma)^{\mathrm{IDM}}=\frac{G_F\alpha^2M_h^3}{128\sqrt{2}\pi^3}\bigg|\underbrace{\frac{4}{3}A_{1/2}\left(\frac{4M_t^2}{M_h^2} \right)+A_1\left(\frac{4M_W^2}{M_h^2} \right)}_{\mathcal{M}^{\mathrm{SM}}}+\underbrace{\frac{\lambda_3 v^2}{2M_{H^\pm}^2}A_0 \left(\frac{4M_{H^\pm}^2}{M_h^2} \right)}_{\delta\mathcal{M}^{\mathrm{IDM}}}\bigg|^2\,, \label{Hloop}
%\end{equation}
\begin{equation}
 \Gamma(h \rightarrow \gamma\gamma)^{\mathrm{IDM}}=\frac{G_F\alpha^2M_h^3}{128\sqrt{2}\pi^3}\bigg|\mathcal{M}^{\mathrm{SM}}+\delta\mathcal{M}^{\mathrm{IDM}}(\m,\lambda_3)\bigg|^2\,, \label{Hloop}
\end{equation}
where $\mathcal{M}^{\textrm{SM}}$ is the SM amplitude and $\delta\mathcal{M}^{\textrm{IDM}}$ is the $H^\pm$ contribution.
The interference between $\mathcal{M}^{\textrm{SM}}$ and $\delta\mathcal{M}^{\textrm{IDM}}$ can be either constructive or destructive. The second source of deviations are possible invisible decays $h\to HH, AA$, which can strongly augment the total decay width $\Gamma^{\textrm{IDM}}(h)$  with respect to the SM case. 
%Partial widths for these decays  are given by:
%\begin{equation}\label{inv-width}
% \Gamma(h\to HH)=\frac{\lambda_{345}^2v^2}{32\pi M_h}\sqrt{1-\frac{4M_{H}^2}{M_h^2}}\, , 
%\end{equation}
%with $M_H$ exchanged to $M_A$ and $\lambda_{345}$ to $\lambda_{345}^-$ ($\lczp^-=\lambda_3+\lambda_4-\lambda_5$), for the $h\to AA$ decay. Using eq.~(\ref{mass}) one can reexpress the couplings $\lc$ and $\lczp^-$ in terms of $M_H,\ M_A,\ \m$ and $\lczp$, and so from eq.~(\ref{rgg}) and~(\ref{inv-width})
%$\rg$   depends only on the masses of the dark scalars and $\lambda_{345}$.
%while there is no dependence on $M_A$ if $M_{A}>M_h/2$
%Those channels are open if $M_{H,A}<M_h/2$,
%\end{itemize}
%We will consider separate cases, when the invisible channels are open or closed, i.e.  regions of  $M_H,\ M_A$  smaller and larger than $M_h/2 \approx  62.5$ GeV. In both cases the precise value of $\rg$ depends on $\m$ and $\lambda_{345}$ (or equivalently $\lc$) and the LHC data can constrain strongly the IDM parameter space, as  will be shown below.
If $h$ can decay invisibly then $\rg$ is always below 1.\cite{Swiezewska:2012eh,Arhrib:2012} For $M_H>M_h/2$ (and $M_A>M_h/2$)  the invisible channels are closed, and $\rg >1$ is possible.
%, with the maximal value of $\rg$ equal to $3.69$ for $M_H = \m = 70$ GeV. 
%If $M_{H} <M_h/2$ then the $h\to HH$ invisible channel is open and it is not possible to obtain $\rg>1$
$\rg$ depends only on the masses of the dark scalars and $\lambda_{345}$ (or $\lambda_3$), so setting a lower bound on $\rg$ leads to upper and lower bounds on $\lczp$  as functions of $M_{H,A,H^\pm}$.\cite{Krawczyk:2013jta}
%We will explore these bounds, as functions of $M_H$ and $\delta_A$ in Sections~\ref{sec-open} and~\ref{sec_Aclosed} for three cases that are in 1$\sigma$ region of the CMS value: $\rg>0.7,\,0.8,\,0.9$, 
%respectively.

\paragraph{$HH, AA$ decay channels open}

%If both $M_H, M_A<M_h/2$ then the LEP constraint for the IDM enforces $\delta_A<8\g$ and so eq.~(\ref{LEPI}) limits the allowed values of the DM particle mass 
%as follows: $M_A+M_H\approx 2 M_H \gtrsim 80$ GeV, so 
%**PS** to bylo troche bez sensu, po po prostu M_H>(M_Z-8)/2~41.
%$M_H>(M_Z - 8 \g)/2 \approx 41\g$.
%**BS** zmieni³am \gtrsim na \approx
 In this region, the invisible decay channels have stronger influence on the value of $\rg$  than the contribution from $H\pm$ loop.\cite{Swiezewska:2012eh}
 % and so the exact value of $\m$ influences the results less than the other scalar masses. 
 %In the following examples we use $\m = 120$~GeV, which is a good benchmark value of the charged scalar mass in the DM analysis for the low and medium DM mass regions, discussed later in section \ref{consequences}. Due to the  dependence of the partial width $\Gamma(h \rightarrow AA)$ on $|\lambda_{345}^-|$ the obtained lower and upper bounds are not symmetric with respect to $\lambda_{345}=0$.
If we demand that $\rg > 0.7$, we get allowed values of $\lambda_{345}$ that are small, typically in range $(-0.04,0.04)$. For $\rg > 0.8$ the allowed values of $\lambda_{345}$ are smaller than for $\rg >0.7$. The condition $\rg > 0.9$ strongly limits the allowed parameter space of the IDM. The allowed  $A,H$ mass difference is $\delta_A \lesssim 2$ GeV, and values of $\lambda_{345}$ are smaller than in the previous cases. Requesting  larger $\rg$ leads to the exclusion of the whole region of masses, apart from $M_H \approx M_A \approx M_h/2$.\cite{Krawczyk:2013jta}

%\paragraph{$\pmb{\textrm{Br}(h\to \textrm{inv})}$} In principle, while discussing the $M_H<M_h/2$ region, one should also include the constraints  from existing LHC data on the invisible channels branching ratio \cite{Djouadi:2012zc,Goudelis:2013uca}. However, constraints on $\lambda_{345}$ obtained by requesting $\textrm{Br}(h\to \textrm{inv}) < 65\, \%$ \cite{ATLAS:2013oma} are up to 50\% weaker than those coming from $\rg$, compare figure \ref{rgg:fig:lam345open}  and figure \ref{inv65}. The limits from the invisible branching ratio start to be comparative with the $\rg$ constraints when $\textrm{Br}(h\to \textrm{inv}) < 20\, \%$, as estimated in \cite{Belanger:2013kya,Dumont:2013mba}.
%

\paragraph{$AA$ decay channel closed}

When the $AA$ decay channel is closed,  the values of $\rg$ do not depend on the value of $M_A$,
%. Since there is only one invisible channel open,  
while the charged scalar contribution becomes more relevant. 
%A clear dependence on the $H^\pm$ mass appears especially for $\m \lesssim 120$ GeV. 
%Figure \ref{Aclosed} shows the limits on $\lambda_{345}$ coupling that allow  values of $\rg$ higher than 0.7, 0.8 and 0.9 for $\m = 70,\,120$ and 500 GeV,  respectively.
%One can see that the larger value of $\rg$ we demand, the smaller values of $\lambda_{345}$ we get. 
If $\rg>0.7$ then an exact value of $\m$ is not crucial for the obtained limits on $\lambda_{345}$, and allowed values of $|\lambda_{345}|$ are of order $ 0.02$. For $\rg >0.8$ the obtained bounds are different for $\m = 70$ GeV and 120 GeV. Smaller $\m$ leads to stronger limits, requiring $|\lambda_{345}|\sim 0.005$, while larger $\m$ allow $|\lambda_{345}| \sim 0.015$. Larger value of $\rg$ leads to  smaller allowed values of $\lambda_{345}$. In the case of $\rg>0.9$ a~large region of DM masses is excluded, as it is not possible to obtain the requested value of $\rg$ for any value of $\lambda_{345}$ if $M_H \lesssim 45 \g$.\cite{Krawczyk:2013jta} 

\paragraph{Invisible decay channels closed}

If $M_A,M_H>M_h/2$,  the invisible channels are closed and the only modification to $\rg$  comes from the charged scalar loop (\ref{Hloop}).
% so the most important parameters are $\m$ and $\lambda_{3}$ (or equivalently $m_{22}^2$). The contribution from the SM ($\mathcal{M}^{\mathrm{SM}}$) is real and negative and $\delta\mathcal{M}^{\mathrm{IDM}}$ is also real with sign correlated with the sign of $\lambda_3$. 
 Enhancement in $\rg$ is possible when $\lambda_3<0$.\cite{Arhrib:2012,Swiezewska:2012eh}
 % with the maximal value of  $\rg$ approached for $\lambda_3=-1.47$.
 % i.e. the smallest value of this parameter allowed by model constraints (\ref{constraints}).
%The contribution to the amplitude from the charged scalar loop ($\delta\mathcal{M}^{\mathrm{IDM}}$) is a decreasing function of $M_{H^\pm}$ so in general the larger $\rg$ is, the smaller $\m$ should be. 
Unitarity and positivity limits on $\lc$ and $\lambda_{345}$ constrain the allowed values of $\m$ and $M_H$ for a given value of $\rg$. For $\rg^\textrm{max}=1.01$ masses of $\m \gtrsim 700$~GeV are excluded, and if $\rg^\textrm{max} =1.02$ this bound is stronger, forbidding $\m \gtrsim 480$~GeV.
% while $\rg>1.2$ gives $70\g<\m<154 \g$ \cite{Swiezewska:2012eh}.
%Since for invisible channels closed $\rg$ depends only on $\m$ and $\lc$ (or $m_{22}^2$), fixing $\rg$ and $\m$ sets the value of $m_{22}^2$. For fixed $m_{22}^2$, $M_H$ depends only on $\lczp$,  eq.~(\ref{mass}). Thus, we can study the correlation between $\m$, $M_H$ and $\lczp$ for different values of $\rg$. 
%Figure \ref{rgg:fig:mhmap} shows the ranges of $\lambda_{345}$ in the $(\m,\delta_{H^\pm})$ plane for two values of $\rg$ close to~1, $\rg = 1.01$ and $1.02$. 
Also, even a small deviation from $\rg =1$ requires a relatively large $\lambda_{345}$, if the mass difference $\delta_{H^\pm}$ is of the order $(50-100)$ GeV. Small values of $|\lambda_{345}|$ are preferred if $\delta_{H^\pm}$ is small.\cite{Krawczyk:2013jta}

%
%
%The blue curve in figure~\ref{rgg:fig:rggmhp} shows the maximal value of $\rg$, which is obtained for maximally allowed negative value of $\lc=-1.47$, as a function of $\m$. In general, as previous studies have shown, the very heavy mass region is consistent with very small deviations from $\rg = 1$, but substantial enhancement of $\rg$ suggested by the central value measured by ATLAS (\ref{rg_atlas}) cannot be reconciled  with this region of masses.
%
%$\rg <1$ is possible if the invisible channels are closed and $\lambda_3 >0$. Requiring that $\rg$ is bounded from above one can also limit the allowed parameter space. For example, if $\rg<0.8$ and invisible channels are closed, then $M_H < 200$ GeV~\cite{Swiezewska:2012eh}. 

\section {Combining  $\rg$  and relic density constraints  on DM \label{consequences}}

Here we compare the limits on the $\lambda_{345}$ parameter  obtained from $\rg$ with those coming from the requirement that the DM relic density is in agreement with the WMAP measurements:    %(\ref{omega}). We use the micrOMEGAs package \cite{Belanger:2013oya} to calculate $\relic$ for chosen values of   DM masses. We demand that the obtained value lies in the 3$\sigma$ WMAP limit:
$0.1018<\Omega_{DM} h^2<0.1234$.
If this condition is fulfilled, then $H$ constitutes 100\% of DM in the Universe. 
Values of 
%$\Omega_{H}h^2>0.1234$ are excluded, while 
$\Omega_H h^2 <0.1018$ are allowed if $H$ is a subdominant DM candidate.

\paragraph{Low DM mass} In the IDM the low DM mass region corresponds to masses of $H$ below 10 GeV, while the other dark scalars are heavier, $M_A \approx M_{H^+} \approx 100$ GeV. 
%In this region the main annihilation channel is $HH \to h \to \bar{b} b$ and 
To obtain the proper relic density, the $HHh$ coupling $(\lambda_{345})$ has to be large, for example $|\lambda_{345}| = (0.35-0.41)$  for CDMS-II favoured mass $M = 8.6$ GeV. %\cite{Agnese:2013rvf} one gets relic density in agreement with bound (\ref{WMAP}) for , while $|\lambda_{345}| \lesssim 0.35$ are excluded. 
The coupling allowed by $\rg \sim 0.7$, i.e. $|\lambda_{345}| \sim 0.02$, is an order of magnitude smaller than needed for $\relic$ and thus we can conclude that the low DM mass region cannot be accommodated in the IDM with recent LHC results.
% irrespective of whether $H$ is the only, or just a subdominant, DM candidate.

%In the low mass region the invisible channel $h \to HH$ is open, meaning that $\rg>1$ is not possible, so we can conclude that $\rg>1$ (\ref{rg_atlas}) excludes the low DM mass region in the IDM. If $\rg<1$, as suggested by the CMS data (\ref{rg_cms}), the low DM mass could be in principle allowed. However, our results, described in the previous section, show, that it is not possible, as the  

%\subsection{Medium DM mass}

\begin{figure}
\begin{minipage}{0.33\linewidth}
\centerline{\includegraphics[width=0.9\linewidth]{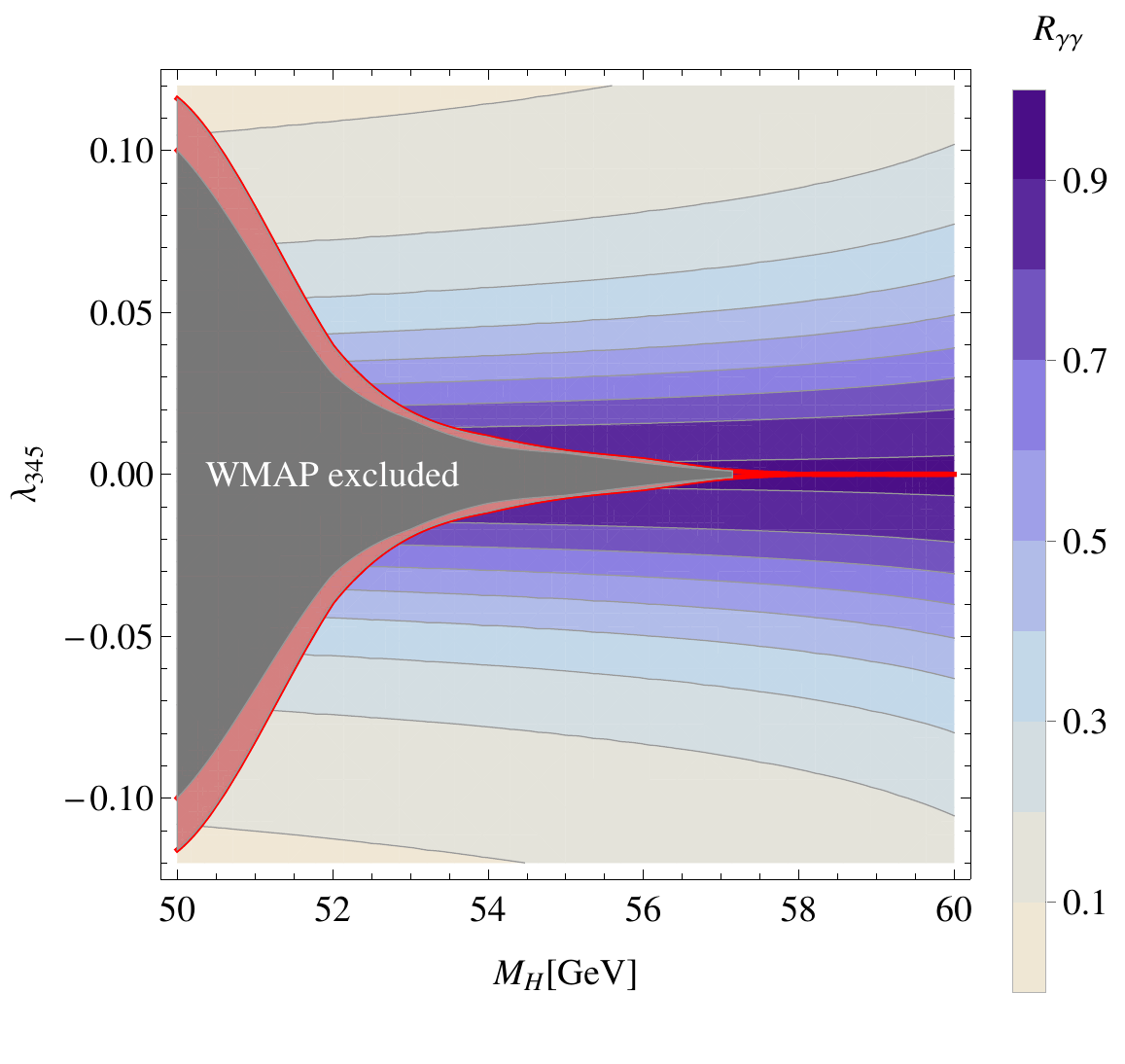}}
\end{minipage}
\hfill
\begin{minipage}{0.32\linewidth}
\centerline{\includegraphics[width=0.9\linewidth]{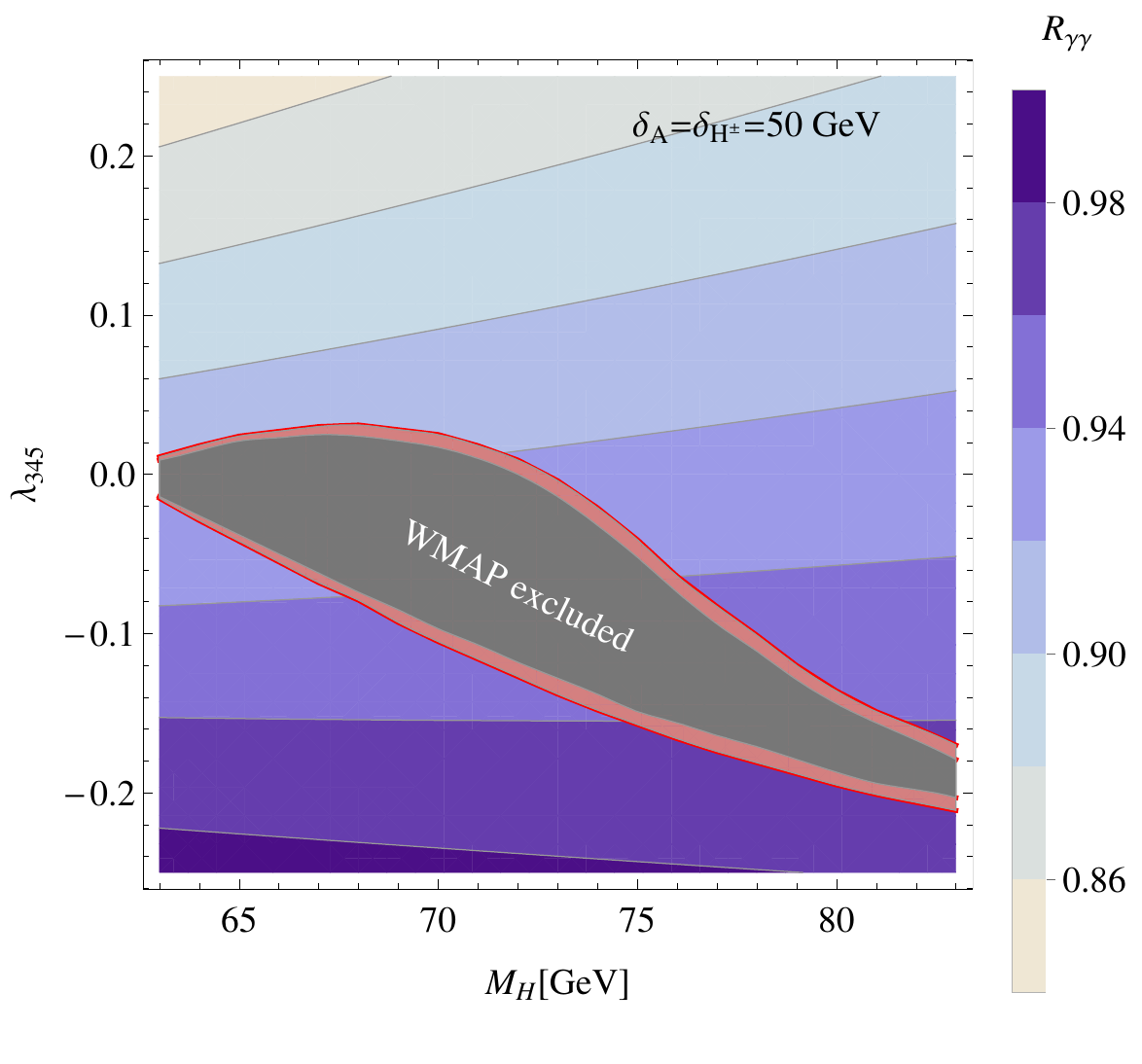}}
\end{minipage}
\hfill
\begin{minipage}{0.32\linewidth}
\centerline{\includegraphics[width=0.9\linewidth]{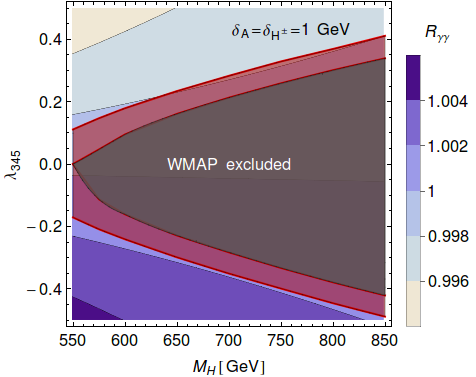}}
\end{minipage}
\caption{Comparison of the values of $\rg$ and region allowed by the relic density measurements for the medium DM mass with (left) $HH$ invisible channel open and $M_A = M_{H^\pm} = 120$ GeV, (central) with $HH$ invisible channel closed and $\delta_A = \delta_{H^\pm} = 50$ GeV and heavy DM mass (right) with $\delta_A = \delta_H^\pm = 1 \g$. Red bound: region in agreement with WMAP. Grey area: excluded by WMAP. $\delta_{A,\pm} = M_{A,H^\pm}-M_H$.}
\label{midOmega}
\end{figure}

%\begin{figure}[t]
%\centering
%\includegraphics[width=.55\textwidth]{RggWMAPopen-eps-converted-to.pdf}
%\caption{Comparison of the values of $\rg$ and region allowed by the relic density measurements for the middle DM mass region with $HH$ invisible channel open and $M_A = M_{H^\pm} = 120$ GeV. Red bound: region in agreement with WMAP (\ref{WMAP}). Grey area: excluded by WMAP. $\rg >0.7$ limits the allowed values of masses to $M_H > 53$ GeV.} \label{midOmega}
%\end{figure}

\paragraph{Medium DM mass: invisible decay channels open} We first consider a case with $M_A = M_{H^\pm} = 120$ GeV and $M_h/2>M_H>50 \g$. 
%In this case the main annihilation channels are $HH \to h \to \bar{f}f$, when the $HHh$ coupling is large enough and $HH \to W^+ W^-$, when the $HHh$ coupling is suppressed, typically leading to $\relic$ above the WMAP limit. Lower values of $M_H$ require rather large $\lambda_{345}$ --- in this sense this region resembles the low DM mass region. As $M_H$ grows towards $M_H = M_h/2$, the value of $\lambda_{345}$ required to obtain the proper relic density gets smaller, leading eventually to the $\relic$ below WMAP limit, apart from extremely tunned and small values of $\lambda_{345}$.
Red bound in the left panel of figure \ref{midOmega} denotes the WMAP-allowed range of $\relic$. 
%Grey excluded region between the WMAP bounds corresponds to $\relic$ too large, leading to the overclosing of the Universe. 
If we consider $H$ as a subdominant DM candidate with $\Omega_H h^2 < \relic$ then also the regions below and above  red bounds in figure \ref{midOmega} are allowed. This usually corresponds to larger values of $\lambda_{345}$. For a large portion of the parameter space limits for $\lambda_{345}$ from $\rg$, even for the least stringent case $\rg >0.7$, cannot be reconciled with the WMAP-allowed region, where $|\lambda_{345}| \sim 0.1$, excluding $M_H\lesssim 53 \g$.

%\begin{figure}[t]
%\centering
%\includegraphics[width=.55\textwidth]{RggWMAPclosed-eps-converted-to.pdf}
%\caption{Comparison of the values of $\rg$ and region allowed by the relic density measurements for the middle DM mass region with $HH$ invisible channel closed and $\delta_A = \delta_{H^\pm} = 50$ GeV. Red bound: region in agreement with WMAP (\ref{WMAP}). Grey area: excluded by WMAP. $\rg >1$ is not possible, unless $H$ is a subdominant DM candidate.} \label{midOmega2}
%\end{figure}

\paragraph{Medium DM mass: invisible decay channels closed} Here we choose $\delta_{H^\pm} = \delta_A = 50$ GeV and $M_H$ varying between $M_h/2$ and $83$ GeV. Figure \ref{midOmega} (central panel) gives the WMAP-allowed range with the corresponding values of $\rg$. The absolute values of $\lambda_{345}$ that lead to the proper relic density are in general larger than in the case of $M_H<M_h/2$. From figure \ref{midOmega} it can be seen that this region of $M_H$ is consistent with $\rg<1$,
%It is in agreement with results obtained before (figure \ref{rgg:fig:mhmap}), as mass difference $\delta_{H^\pm} = 50$ GeV and $\rg>1$ requires $\lambda_{345} \lesssim -0.3$, a value smaller than the one obtained from the relic density limits.
and that $\rg>1$ and $\Omega_{DM} h^2$ constraints  cannot be fulfilled for the middle DM mass region. If the IDM is the source of all DM in the Universe and $M_H \approx(63-83)$ GeV then the maximal value of $\rg$ is around $0.98$. A~subdominant DM candidate, which corresponds to larger $\lambda_{345}$, is consistent with $\rg>1$. 

\paragraph{Heavy DM mass: almost degenerated particle spectra} In this case it is possible to get $\rg>1$ and be with agreement with WMAP, as shown in right panel in figure \ref{midOmega} for $M_H \gtrsim 500$ GeV and $\delta_A = \delta_\pm = 1 \g$, although deviation from $\rg =1$ is very small.

\section{Summary}

The DM candidate from the IDM is consistent with the WMAP results on the DM relic density and in three regions of masses it can explain 100~\% of the DM in the Universe. In a large part of the parameter space it can also be considered as a subdominant DM candidate.
Measurements of the diphoton ratio $\rg$ done at the LHC set strong limits on masses of the DM and other dark scalars, and their self-couplings. 

We can exclude \textit{the low DM mass region} in the IDM, i.e. $M_H \lesssim 10$ GeV, as values of $|\lambda_{345}|$ needed for the proper $\Omega_{DM} h^2$ 
are an order of magnitude larger than those allowed by assuming that $\rg>0.7$. In \textit{the medium mass region} $\rg>1$ favours degenerated $H$ and $H^\pm$. When the mass difference is large, $\delta_{H^\pm} \approx 50$ GeV, then values of $|\lambda_{345}|$ that provide $\rg>1$ are bigger than those allowed by WMAP. We conclude it is not possible to have $\rg >1$ and all DM in the Universe explained by the IDM in the medium DM mass
region. If $\rg>1$ then $H$ may be a subdominant DM candidate. If $\rg <1$ then $M_H\approx (63-80)$ GeV can explain 100\% of DM in the Universe.
For \textit{heavy DM particles} it is possible to obtain $\rg>1$ and fulfill WMAP bounds, although deviation from $\rg =1$ is  small.

\section*{Acknowledgments}

This work was supported in part by the grant  NCN OPUS 2012/05/B/ST2/03306 (2012-2016).

%\section*{Appendix}
%
% We can insert an appendix here and place equations so that they are
%given numbers such as Eq.~\ref{eq:app}.
%\be
%x = y.
%\label{eq:app}
%\ee


\section*{References}

\begin{thebibliography}{99}
%\bibitem{ja}C Jarlskog in {\em CP Violation}, ed. C Jarlskog
%(World Scientific, Singapore, 1988).
%
%\bibitem{ma}L. Maiani, \Journal{\PLB}{62}{183}{1976}.
%
%\bibitem{bu}J.D. Bjorken and I. Dunietz, \Journal{\PRD}{36}{2109}{1987}.
%
%\bibitem{bd}C.D. Buchanan {\it et al}, \Journal{\PRD}{45}{4088}{1992}.

\bibitem{Cao:2007rm}
Q.-H. Cao {\it et al},
%``{Observing the Dark Scalar Doublet and
%  its Impact on the Standard-Model Higgs Boson at Colliders},''
   {\em  Phys.Rev.}, vol.~D76, p.~095011, 2007, arXiv:0708.2939.

\bibitem{Barbieri:2006dq}
R.~Barbieri {\it et al},
%``Improved naturalness with a heavy
%  higgs: An alternative road to lhc physics,'' 
  {\em Phys. Rev.}, vol.~D74,
  p.~015007, 2006, hep-ph/0603188.

\bibitem{Dolle:2009fn}
E.~M. Dolle and S.~Su, 
%``{The Inert Dark Matter},'' 
{\em Phys.Rev.}, vol.~D80,
  p.~055012, 2009, arXiv:0906.1609.

\bibitem{Bergstrom:2012fi}
L.~Bergstrom, 
%``{Dark Matter Evidence, Particle Physics Candidates and
%  Detection Methods},'' 
  2012, arXiv:1205.4882.  

\bibitem{Krawczyk:2010}
I.~Ginzburg {\it et al},
%``{Evolution of
%  Universe to the present inert phase},'' 
  {\em Phys.Rev.}, vol.~D82, p.~123533,
  2010, arXiv:1009.4593.

\bibitem{Sokolowska:2011aa}
D.~Sokolowska, 
%``{Dark Matter Data and Quartic Self-Couplings in Inert Doublet
%  Model},'' 
  {\em Acta Phys.Polon.}, vol.~B42, p.~2237, 2011, arXiv:1112.2953.

%\bibitem{Sokolowska:2011sb}
%D.~Sokolowska, ``{Dark Matter Data and Constraints on Quartic Couplings in
%  IDM},'' 2011, 1107.1991.
%
%%\cite{Sokolowska:2011yi}
%\bibitem{Sokolowska:2011yi} 
%  D.~Sokolowska,
%  ``Temperature evolution of physical parameters in the Inert Doublet Model,''
%  arXiv:1104.3326 [hep-ph].
%  %%CITATION = ARXIV:1104.3326;%%
%  %4 citations counted in INSPIRE as of 10 Jun 2013

\bibitem{Kanemura:1993}
S.~Kanemura {\it et al},
%``{Lee-Quigg-Thacker bounds for Higgs
%  boson masses in a two doublet model},'' 
  {\em Phys.Lett.}, vol.~B313,
  pp.~155--160, 1993, hep-ph/9303263.

%\bibitem{Akeroyd:2000}
%A.~G. Akeroyd {\it et al},
%%``{Note on tree level unitarity in
%%  the general two Higgs doublet model},'' 
%  {\em Phys.Lett.}, vol.~B490,
%  pp.~119--124, 2000, hep-ph/0006035.

\bibitem{Swiezewska:2012}
B.~Swiezewska, 
%``{Yukawa independent constraints for 2HDMs with a 125 GeV Higgs
%  boson},'' 
  2012, arXiv:1209.5725.

\bibitem{Gustafsson:2009}
E.~Lundstrom {\it et al},
%``{The Inert Doublet Model and LEP
%  II Limits},''
   {\em Phys.Rev.}, vol.~D79, p.~035013, 2009, arXiv:0810.3924.

\bibitem{ATLAS:2013oma}
%``{Measurements of the properties of the Higgs-like boson in the two photon
%  decay channel with the ATLAS detector using 25 $\mathrm{fb}^{-1}$ of
%  proton-proton collision data},'' 2013.
ATLAS Collaboration, ATLAS-CONF-2013-012.

\bibitem{CMStalk}
C.~Mariotti, 
%``Recent results on higgs studies at {CMS},'' 2013.
\newblock Slides of a talk given at CERN, 15th of April, 2013.

\bibitem{Djouadi:2005}
A.~Djouadi, {\em Phys.Rept.}, vol.~459,
  pp.~1--241, 2008; {\em Phys.Rept.}, vol.~457, pp.~1--216, 2008.


%\bibitem{Djouadi:2005}
%A.~Djouadi,
%% ``{The Anatomy of electro-weak symmetry breaking. II. The Higgs
%%  bosons in the minimal supersymmetric model},''
%   {\em Phys.Rept.}, vol.~459,
%  pp.~1--241, 2008, hep-ph/0503173.
%
%
%\bibitem{Djouadi:2005sm}
%A.~Djouadi, 
%%``{The Anatomy of electro-weak symmetry breaking. I: The Higgs
%%  boson in the standard model},'' 
%  {\em Phys.Rept.}, vol.~457, pp.~1--216, 2008,
%  hep-ph/0503172.

%\bibitem{Ma:2007}
%Q.-H. Cao {\it et al},
%% ``{Observing the Dark Scalar Doublet and
%%  its Impact on the Standard-Model Higgs Boson at Colliders},'' 
%{\em  Phys.Rev.}, vol.~D76, p.~095011, 2007, 0708.2939.
%
%\bibitem{Posch:2010}
%P.~Posch, 
%%``{Enhancement of $h \to \gamma \gamma$ in the Two Higgs Doublet
%%  Model Type I},'' 
%  {\em Phys.Lett.}, vol.~B696, pp.~447--453, 2011, 1001.1759.

\bibitem{Arhrib:2012}
A.~Arhrib {\it et al},
%``{$H\to \gamma \gamma$ in Inert Higgs  Doublet Model},'' 
  {\em Phys.Rev.}, vol.~D85, p.~095021, 2012.

\bibitem{Swiezewska:2012eh}
B.~Swiezewska and M.~Krawczyk, 
%``{Diphoton rate in the Inert Doublet Model with
%  a 125 GeV Higgs boson},'' 
 {\em Phys. Rev.}, vol.~D88, p.~035019, 2013,
arXiv:1212.4100.
  
  %\cite{Krawczyk:2013jta}
\bibitem{Krawczyk:2013jta}
  M.~Krawczyk {\it et al},
  %``Constraining Inert Dark Matter by R_{\gamma\gamma} and WMAP data,'' 
  %%CITATION = ARXIV:1305.6266;%%
  JHEP {\bf 09} (2013) 055, 2013, 
  arXiv:1305.6266 [hep-ph].
  %2 citations counted in INSPIRE as of 21 Aug 2013


\end{thebibliography}
\end{document}